\documentclass[prl,nofootinbib,preprint,superscriptaddress,twocolumn,10pt]{revtex4}
\pdfoutput=1
\usepackage[T1]{fontenc}
\usepackage{amsmath,amssymb}
\usepackage{epsfig}
\usepackage{float}
\usepackage{graphicx}
\usepackage[usenames,dvipsnames]{color}
\usepackage{subfigure}
\usepackage{slashed}
\usepackage[colorlinks,citecolor=blue]{hyperref}
\usepackage{color}
\usepackage{cancel}
\usepackage{changes}
\usepackage{comment}

\begin{document}

	\title{Higgs Quadruplet Impact on $W$ Mass Shift, Dark Matter, and LHC Signatures}	

	\author{Talal Ahmed Chowdhury}
	\affiliation{Department of Physics, University of Dhaka, P.O. Box 1000, Dhaka, Bangladesh}
	\author{Kareem Ezzat}
		\affiliation{Department of Mathematics, Faculty of Science, Ain Shams University, Cairo 11566, Egypt}
		
	\affiliation{Center for Fundamental Physics, Zewail City of Science and Technology, Sheikh Zayed,12588, Giza, Egypt}
	\author{Shaaban Khalil}
	\affiliation{Center for Fundamental Physics, Zewail City of Science and Technology, Sheikh Zayed,12588, Giza, Egypt}
	
		\author{Ernest Ma}
	\affiliation{Department of Physics and Astronomy, University of California, Riverside, California 92521, USA}
	\author{Dibyendu Nanda}
	\affiliation{School of Physics, Korea Institute for Advanced Study, Seoul 02455, Republic of Korea}

\begin{abstract}
The addition of a Higgs quadruplet to the standard model (SM) of quarks and leptons would shift the $W$ boson mass upward. It could also facilitate the production of dark matter through the conventional thermal freeze-out scenario via Yukawa interaction with the Higgs quadruplet or freeze-in production from the decay of SM Higgs. We investigate the same-sign lepton smoking gun signature of the double-charged scalar component of the Quadruplet Higgs at the LHC.
\end{abstract}
\maketitle
The Standard Model (SM) of quarks and leptons has one Higgs doublet $\Phi=(\phi^+,\phi^0)$.  When the gauge symmetry $SU(2)_L \times U(1)_Y$ undergoes spontaneous symmetry breaking due to $\langle \phi^0 \rangle=v_0$, a well-established tree-level condition emerges: $\rho=M^2_W/M^2_Z \cos^2 \theta_W=1$.  The recent measurement of the $W$ boson's mass by the CDF experiment is given by $m_W^{\rm CDF} = 80.4335 \pm 0.094 ~ {\rm GeV}$ \cite{CDF:2022hxs}, which shows $7 \sigma$ deviation from the SM prediction: $m_W^{\rm SM} = 80.357 \pm 0.006$ GeV. This has prompted theorists to speculate about potential new physics contributing to this unexpected increase in the $W$ boson's mass.

If an additional Higgs multiplet with an isospin $I$ is introduced to the SM, the vacuum expectation value $v_1$ of its neutral component, having third component of isospin $I_3$, contributes $2(g^2/\cos^2 \theta_W) I_3^2 v_1^2$ to $M_Z^2$, and $g^2 [I(I+1)-I_3^2] v_1^2$ to $M_W^2$.   As a result, $\rho$ deviates from unity except for certain specific values of $I$ and $I_3$. Notably, the $\rho$ parameter derived from electroweak global fits is $\rho =1.0002 \pm 0.0009$ \cite{ParticleDataGroup:2022pth}. Therefore, this leads to stringent constraints on the value of $v_1$.

In this letter, we study the effect of adding a Higgs quadruplet $\zeta = (\zeta^{++},\zeta^+,\zeta^0,\zeta^-)$ to the SM.  The motivation for extending the SM with a Higgs quadruplet primarily stems from the quest to resolve the neutrino mass problem. 
The contribution of $\langle \zeta^0 \rangle = v_\zeta$  
to $M_W^2$ is $7 g^2 v_\zeta^2/2$, and that to $M_Z^2$ is 
$g^2 v_\zeta^2/2\cos^2 \theta_W$.  This could then explain the  
$W$ mass shift, in agreement with the recent precision 
measurement~\cite{CDF:2022hxs}.  Whereas $\zeta$ is necessary for the quintuplet neutrino seesaw 
mechanism~\cite{Liao:2010cc,Kumericki:2012bh}, it may also be the connector in Type III 
seesaw~\cite{Foot:1988aq}.  In our model, another use of $\zeta$ is 
proposed, as the connector to the dark sector consisting of a  
neutral Majorana fermion singlet $N$ and a Dirac fermion quadruplet 
$\Sigma=(\Sigma^{++},\Sigma^+,\Sigma^0,\Sigma^-)$.

Consider the Higgs quadruplet $\zeta$.  It interacts with 
$(W^+,W^0,W^-)$ and the $U(1)_Y$ gauge boson $B$ according to  
$|(\partial_\mu - i g {\cal W}^{(4)}_\mu - i (g'/2) B_\mu) \zeta|^2$, where 
\begin{equation}
{\cal W}^{(4)} = 
\begin{pmatrix}
(3/2)W^0 & \sqrt{3/2}W^+ & 0 & 0 \cr \sqrt{3/2}W^- & 
(1/2)W^0 & \sqrt{2}W^+ & 0 \cr 0 & \sqrt{2}W^- & -(1/2)W^0 & \sqrt{3/2}W^+ 
\cr 0 & 0 & \sqrt{3/2}W^- & -(3/2)W^0
\end{pmatrix}.
\end{equation}
For the fermion quadruplet $\Sigma$, the corresponding interaction is 
$\overline{\Sigma} \gamma^\mu (\partial_\mu - i g {\cal W}^{(4)}_\mu - 
i (g'/2) B_\mu) \Sigma$.  
We assume that $\Sigma$ is odd under a dark $Z_2$ symmetry together with 
a neutral singlet fermion $N$.  Hence $\zeta^\dagger \Sigma N$ is allowed 
and $\zeta$ becomes the connection between the SM and the dark sector. The Higgs potential consisting of $\Phi$ and $\zeta$ is given by~\cite{Kumericki:2012bh} 

\begin{widetext}
 \begin{eqnarray} 
 V &= -\mu_1^2 \Phi^\dagger \Phi + \mu_2^2 \zeta^\dagger \zeta + {1 \over 2} 
\lambda_1 (\Phi^\dagger \Phi)^2 + {1 \over 2} \lambda_2 
(\zeta^\dagger \zeta)^2+ {1 \over 2} \lambda'_2 [\zeta^\dagger \zeta]_i 
[\zeta^\dagger \zeta]_i + \lambda_3 (\Phi^\dagger \Phi)
(\zeta^\dagger \zeta) + \lambda'_3 [\Phi^\dagger \Phi]_i 
[\zeta^\dagger \zeta]_i + \nonumber \\  
& \left\{ {1 \over 2} \lambda''_3 
[\Phi^\dagger \zeta]_i [\Phi^\dagger \zeta]_i + H.c. \right\}   
- \left\{ {1 \over 3} \lambda_4 [\zeta^\dagger \Phi]_i 
[\Phi^\dagger \Phi]_i + H.c. \right\} - \left\{ {1 \over 3} \lambda_5 
[\Phi^\dagger \zeta^\dagger]_i [\zeta \zeta]_i + H.c. \right\},
\label{potential}
\end{eqnarray}
\end{widetext}

where the sum over $i=1,2,3$ represents the components of the tensor product of the scalar fields. There are in principle 4 quartic terms of the type $(\zeta^\dagger \zeta)(\zeta^\dagger \zeta)$, corresponding to the pairings $1 \times 1$, $3 \times 3$, $5 \times 5$, and $7 \times 7$, where  $1 \times 1$ indicates the multiplication between singlet components of each product, whereas $3 \times 3$ represents the multiplication between triplet components, and so on. However, the $5 \times 5$ term is proportional to $1 \times 1$.    The sum of $3 \times 3$ and $7 \times 7$ is also proportional to $1 \times 1$.  Hence there are only two independent terms.  
As for the terms mixing $\zeta$ with $\Phi$,  $(\Phi^\dagger \Phi)(\zeta^\dagger \zeta)$ has two terms, whereas $(\Phi^\dagger \zeta)(\Phi^\dagger \zeta)$, $(\zeta^\dagger \Phi)(\Phi^\dagger \Phi)$, $(\Phi^\dagger \zeta^\dagger)(\zeta \zeta)$, each has just one term. Now, the different terms of the scalar potential can be expressed by using the tensorial notation of $\zeta$ as follows
\begin{widetext}
\begin{eqnarray}
[\zeta^\dagger \zeta]_i[\zeta^\dagger \zeta]_i &=& [3|\zeta^{++}|^2 + 
|\zeta^+|^2 - |\zeta^0|^2 - 3|\zeta^-|^2]^2 \nonumber \\ 
&+& 4|\sqrt{3} \zeta^{++} \overline{\zeta^+} + 2 \zeta^+ \overline{\zeta^0} 
+ \sqrt{3} \zeta^0 \overline{\zeta^-}|^2, \\~
[\Phi^\dagger \Phi]_i [\zeta^\dagger \zeta]_i &=& [|\phi^+|^2 - |\phi^0|^2] 
[3|\zeta^{++}|^2 + |\zeta^+|^2 - |\zeta^0|^2 - 3|\zeta^-|^2] \nonumber \\ 
&+& \{ 2 \phi^+ \overline{\phi^0} [\sqrt{3} \zeta^+ \overline{\zeta^{++}} + 
2 \zeta^0 \overline{\zeta^+} + \sqrt{3} \zeta^- \overline{\zeta^0}] + H.c. \}, 
\\~ 
[\Phi^\dagger \zeta]_i [\Phi^\dagger \zeta]_i &=& (\overline{\phi^0} \zeta^0 
+ \phi^- \zeta^+)^2 - (\overline{\phi^0} \zeta^+ + \sqrt{3} \phi^- \zeta^{++}) 
(\sqrt{3} \overline{\phi^0} \zeta^- + \phi^- \zeta^0),  \\~ 
[\zeta^\dagger \Phi]_i [\Phi^\dagger \Phi]_i &=& [\overline{\zeta^0} \phi^0 + \overline{\zeta^+} \phi^+][|\phi^0|^2 - |\phi^+|^2] + [\overline{\zeta^+} 
\phi^0 + \sqrt{3} \overline{\zeta^{++}} \phi^+] \phi^+ \overline{\phi^0} 
\nonumber \\ &-& [\sqrt{3}\overline{\zeta^-} \phi^0 + \overline{\zeta^0} 
\phi^+] \phi^0 \phi^-,\\~
[\Phi^\dagger \zeta^\dagger]_i [\zeta \zeta]_i &=& (\overline{\phi^0} 
\overline{\zeta^0} - \sqrt{3} \phi^- \overline{\zeta^-}) (\zeta^0 \zeta^0 - 
\sqrt{3} \zeta^+ \zeta^-) + (\overline{\phi^0} \overline{\zeta^+} - \phi^- 
\overline{\zeta^0}) (\zeta^+ \zeta^0 - 3 \zeta^{++} \zeta^-) \nonumber \\ 
&+& (\sqrt{3} \overline{\phi^0} \overline{\zeta^{++}} - \phi^- 
\overline{\zeta^+}) (\zeta^+ \zeta^+ - \sqrt{3} \zeta^{++} \zeta^0). 
\end{eqnarray}
\end{widetext}

Let $\langle \phi^0 \rangle = v_0$ and $\langle \zeta^0 \rangle = v_\zeta$,  then the minimum of $V$ is determined by 
  $0 = v_0[-\mu_1^2 \!+\! \lambda_1 v_0^2 \!+\! (\lambda_3 \!+\! \lambda'_3 \!+\! \lambda''_3) 
v_\zeta^2 \!-\! \lambda_4 v_0 v_\zeta] \!-\! {1 \over 3} \lambda_5 v_\zeta^3$ and 
$0 = v_\zeta[\mu_2^2 \!+\! (\lambda_2 \!+\! \lambda'_2) v_\zeta^2 \!+\! (\lambda_3 + 
\lambda'_3 \!+\! \lambda''_3) v_0^2 \!-\! \lambda_5 v_0 v_\zeta] \!-\! {1 \over 3} 
\lambda_4 v_0^3$. For positive and large $\mu_2^2$, 
$v_\zeta \simeq \lambda_4 v_0^3/3\mu_2^2 << v_0$. 
This is the analog of the scalar seesaw studied previously~\cite{Ma:1998dn,Ma:2000cc}. 
Let $h=\sqrt{2}Re(\phi^0)$ and $H=\sqrt{2}Re(\zeta^0)$, then their 
$2 \times 2$ mass-squared matrix is

\begin{widetext}
\begin{equation}
{\cal M}^2_{hH} = \begin{pmatrix}
 2\lambda_1 v_0^2 - \lambda_4 v_0 v_\zeta + 
\lambda_5 v_\zeta^3/3v_0 & - \lambda_4 v_0^2 + 2 (\lambda_3 + \lambda'_3 + 
\lambda''_3) v_\zeta v_0 -  \lambda_5 v_\zeta^2 \cr - \lambda_4 v_0^2 + 
2 (\lambda_3 + \lambda'_3 + \lambda''_3) v_\zeta v_0 -  \lambda_5 v_\zeta^2 & 
\lambda_4 v_0^3/3v_\zeta - \lambda_5 v_0 v_\zeta + 
2 (\lambda_2 + \lambda'_2) v_\zeta^2
\end{pmatrix}.
\end{equation}    
\end{widetext}

The linear combination proportional to $v_0 Im(\phi^0) + v_\zeta Im(\zeta^0)$ 
becomes the Goldstone boson for $Z$ whereas its orthogonal combination has 
mass-squared $= (\lambda_4 v_0/3v_\zeta + \lambda_5 v_\zeta/3v_0) (v_0^2 + v_\zeta^2)$. The $3 \times 3$ mass-squared matrix spanning 
$(\phi^+,\zeta^+,\overline{\zeta^-})$ is
\begin{widetext}
\begin{eqnarray}
{\cal M}^2_{+} &=& \begin{pmatrix}-2\lambda'_3 v_\zeta^2 + 7\lambda_4 v_0 v_\zeta/3 
& 4 \lambda'_3 v_0 v_\zeta - 2 \lambda_4 v_0^2/3 & 2\sqrt{3} \lambda'_3 
v_0 v_\zeta + \lambda_4 v_0^2/\sqrt{3} \cr 4 \lambda'_3 v_0 v_\zeta - 2 
\lambda_4 v_0^2/3 & 6 \lambda'_2 v_\zeta^2 -2 \lambda'_3 v_0^2 
+ \lambda_4 v_0^3/3v_\zeta & 4\sqrt{3} \lambda'_2 v_\zeta^2 \cr 
2\sqrt{3} \lambda'_3 v_0 v_\zeta + \lambda_4 v_0^2/\sqrt{3} & 
4\sqrt{3} \lambda'_2 v_\zeta^2 & 8 \lambda'_2 v_\zeta^2 + 2 \lambda'_3 v_0^2 
+ \lambda_4 v_0^3/3v_\zeta
\end{pmatrix}\nonumber \\
&+& \begin{pmatrix}-\lambda''_3 v_\zeta^2 + \lambda_5 v_\zeta^3/3v_0 & \lambda''_3 
v_\zeta v_0/2 + \lambda_5 v_\zeta^2/3 & \lambda_5 v_\zeta^2/\sqrt{3} \cr 
\lambda''_3 v_\zeta v_0/2 + \lambda_5 v_\zeta^2/3 & -\lambda''_3 v_0^2 + 
\lambda_5 v_\zeta v_0/3 & -\sqrt{3} \lambda''_3 v_0^2/2 + \lambda_5 v_\zeta 
v_0/\sqrt{3} \cr \lambda_5 v_\zeta^2/\sqrt{3} & -\sqrt{3} \lambda''_3 v_0^2/2 
+ \lambda_5 v_\zeta v_0/\sqrt{3} & -\lambda''_3 v_0^2 + \lambda_5 v_\zeta v_0
\end{pmatrix}.
\end{eqnarray}
\end{widetext}

The linear combination proportional to 
$v_0 \phi^+ + 2 v_\zeta \zeta^+ - \sqrt{3} v_\zeta \overline{\zeta^-}$ 
becomes the Goldstone boson for $W^+$ whereas the other two orthogonal 
combinations have mass-squared 
$=\lambda_4 v_0^3/3v_\zeta - \lambda''_3 v_0^2 \pm \sqrt{4(\lambda'_3 v_0^2)^2 + 3(\lambda''_3 v_0^2)^2/4}$, 
neglecting terms of order $v_0 v_\zeta$ and smaller.  The doubly charged 
scalar boson $\zeta^{++}$ has mass-squared 
$= \lambda_4 v_0^3/3v_\zeta - 4 \lambda'_3 v_0^2 - \lambda''_3 v_0^2 + 
3 \lambda_5 v_0 v_\zeta - 4 \lambda'_2 v_\zeta^2$. The other relevant renormalizable interactions can be written as

\begin{widetext}
\begin{equation}
    \mathcal{L_{\rm Yuk}}= i \overline{\Sigma} \overline{\sigma}^\mu D_{\mu} \Sigma + i\overline{N}\gamma^\mu \partial_{\mu} N + \left[ y_N \zeta^\dagger \overline{N}\Sigma + M_{\Sigma}\overline{\Sigma}\Sigma + M_{N} \overline{N} N + H.c.\right].
    \label{lag:tot}
\end{equation}    
\end{widetext}

Let us now discuss the impact of our model in light of the CDF collaboration measurement of the W boson mass $M_W = 80433.5 \pm 9.4$ MeV \cite{CDF:2022hxs} collected at the CDF-II detector of Fermilab Tevatron collider. The recently measured value of $W-$mass has a 7$\sigma$ departure from the SM expectation ($M_{W}=80357\pm6$ MeV). This has led to various different proposals on the feasible implications and interpretations related to electroweak precision parameters \cite{deBlas:2022hdk, Strumia:2022qkt, Asadi:2022xiy, Lu:2022bgw}, BSM physics like DM \cite{Fan:2022dck, Zhu:2022tpr, Zhu:2022scj,Kawamura:2022uft, Nagao:2022oin, Liu:2022jdq}, additional scalar fields \cite{Chen:2022ocr, Sakurai:2022hwh, Cacciapaglia:2022xih,Song:2022xts, Bahl:2022xzi,Cheng:2022jyi,Babu:2022pdn,Heo:2022dey,Ahn:2022xax,Zheng:2022irz,FileviezPerez:2022lxp,Kanemura:2022ahw}, effective field theory \cite{Fan:2022yly, Bagnaschi:2022whn}, supersymmetry \cite{Du:2022pbp, Tang:2022pxh,Yang:2022gvz, Athron:2022isz,Ghoshal:2022vzo} and several others \cite{Yuan:2022cpw, Athron:2022qpo, Blennow:2022yfm, Heckman:2022the,Lee:2022nqz, DiLuzio:2022xns,Paul:2022dds,Biekotter:2022abc,Balkin:2022glu,Cheung:2022zsb,Du:2022brr, Endo:2022kiw, Crivellin:2022fdf, Arias-Aragon:2022ats}. We study the fit to the new measurement ~\cite{CDF:2022hxs} of the $W$ boson mass,  using newly added quadruplet $\zeta$. The new physics contributions to $W-$boson mass anomaly can be parametrised in terms of the oblique parameters S, T, U \cite{Peskin:1990zt, Peskin:1991sw}. Considering the U parameter to be vanquished, any BSM physics contribution to $W-$ boson mass can be parametrised in terms of $S$ and $T$ parameters. Taking the fine-structure constant $\alpha$, the Fermi constant $G_F$, and Z boson mass $M_Z$ as input parameters, the fitting of $S, T$ parameters in view of the recent $W$-mass anomaly has been discussed in \cite{Bagnaschi:2022whn}. It is very important to note that, a change in the oblique parameters due to BSM physics will also change the precisely measured weak mixing angle $\theta_W$. The $\sin^2_{\theta_W}(m_Z)_{\overline{MS}}$ and the mass of W boson $m_W$ can be expressed in terms of S and T parameters as \cite{Kumar:2013yoa},
\begin{align}
    M_{W}&= 80.357\text{ GeV} \left(1-0.0036\,S+0.0056\,T \right), \nonumber \\
\sin^2_{\theta_W}(m_Z)_{\overline{\rm MS}}&= 0.23124 \left(1-0.0157\,S+0.0112\,T \right). 
\label{TS}
\end{align}

From the above two equations, it can be inferred that the compatibility of the new measurement of $W-$boson mass with the $\theta_W$ requires both S and T parameters to be non-zero, also seen from the fits shown in \cite{Bagnaschi:2022whn}.
However, in the case of a scalar quadruplet $\zeta$ with hypercharge $1/2$, we will get a correction to the T parameter only as also shown in \cite{Strumia:2022qkt}. To take into account the enhanced $W-$ boson mass, the required limit on the T parameter is
$T = 0.17\pm 0.020889$.

\begin{figure}[t!]
    \includegraphics[scale=0.24]{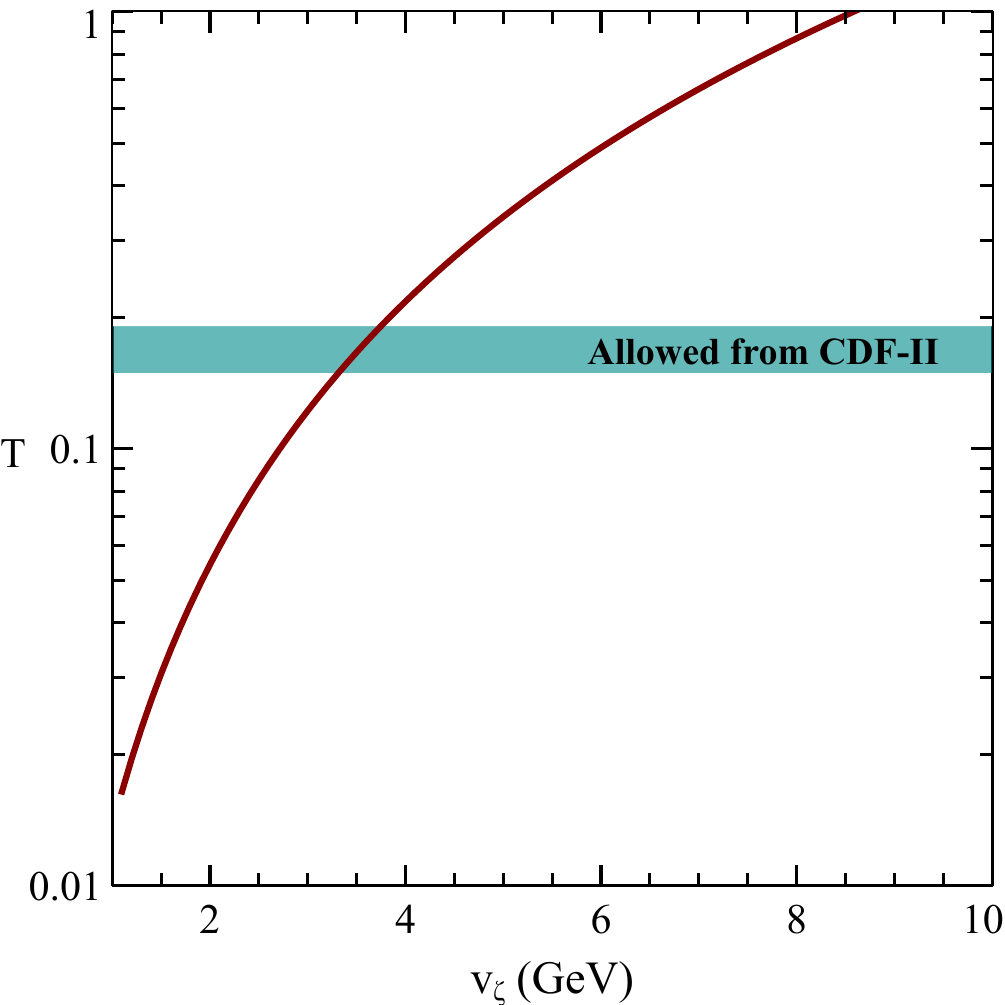}~~\includegraphics[scale=0.24]{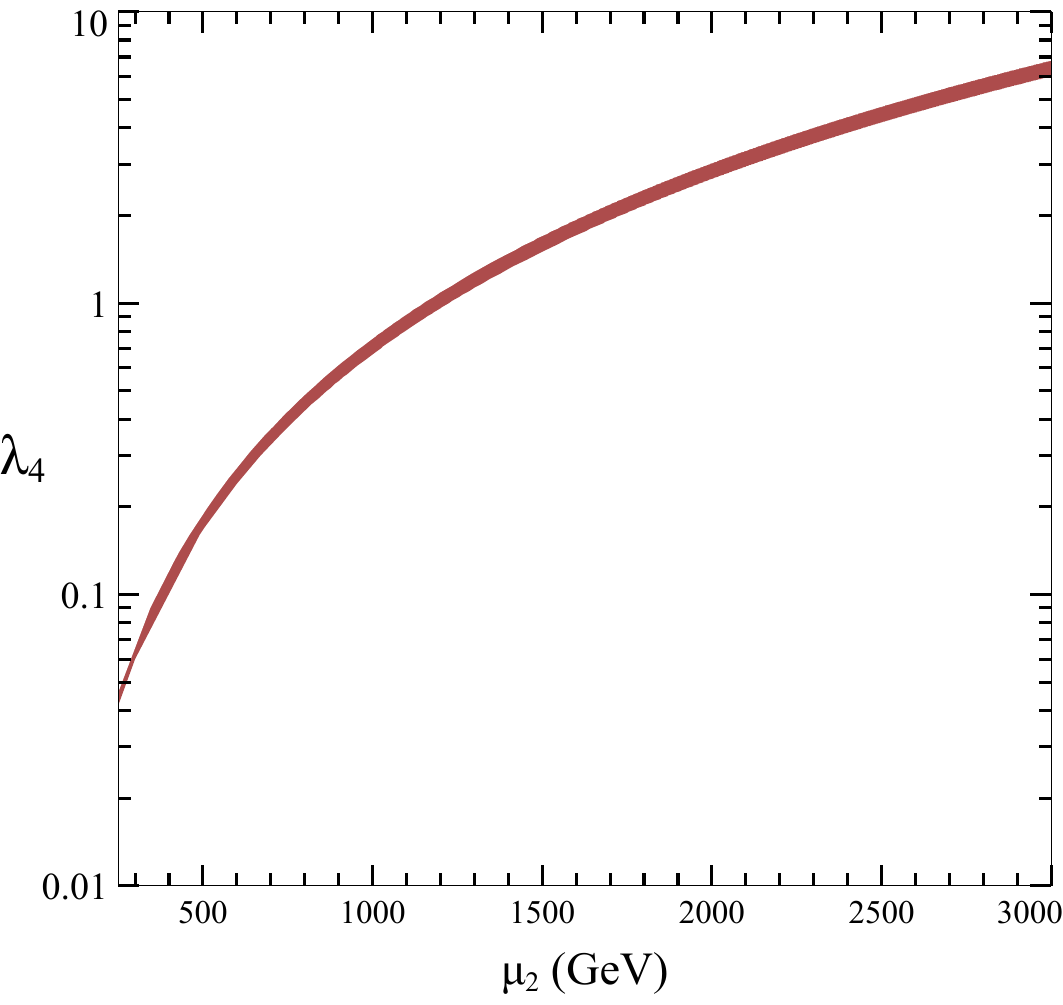}
    \caption{The allowed values of $v_\zeta$ and the parameter space in the $\mu_2 - \lambda_4$ plane from the $M_W$ anomaly.}
    \label{fig:wmass}
\end{figure}

However, following Eq. \eqref{TS}, any change in the T parameter with $S=0$ will put the weak mixing angle $\sin^2{\theta_W}$ in tension with the LEP data. The above-mentioned range of T would imply that $\sin^2{\theta_W}$ should lie in between $0.230746-0.230854$. An additional contribution to $S, T$ parameters would be required to reduce this tension. The $T$ parameter will get a new contribution at tree-level from the vev of $\zeta\,\left(v_\zeta\right)$ and can be written as $T={6 v_\zeta^2}/{\alpha v^2}$ \cite{Dawson:2017vgm, Murphy:2020rsh, Strumia:2022qkt}. This can fit the $M_W$ anomaly for $v_\zeta\approx 3 $ GeV as shown in the left panel of the figure \ref{fig:wmass}. As $v_\zeta$ can be presented in terms of the bare mass term of $\zeta$ ($\mu_2$) and the quartic coupling $\lambda_4$ as they are related via $\mu_2\approx\left({\lambda_4 v^3}/{3 v_\zeta}\right)^{1/2}$, the corresponding constraints on $v_\zeta$ can also be translated in the $\mu_2 - \lambda_4$ plane, as shown in the right panel of the figure \ref{fig:wmass}. Importantly, $\lambda_{4}$ will also decide the mass of the doubly charged scalar $\zeta^{++}$ and we have discussed the impact of $\zeta^{++}$ in collider in the later part of this letter. 

\begin{figure}
    \centering
    \includegraphics[scale=0.25]{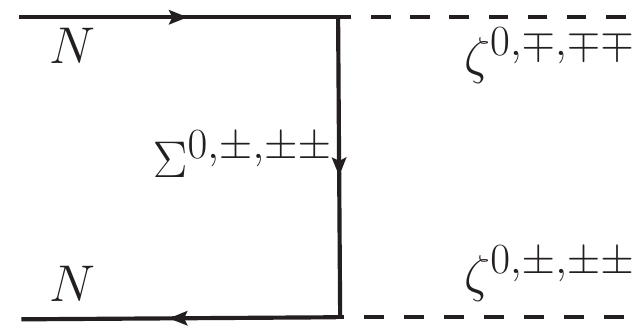}\,
        \includegraphics[scale=0.25]{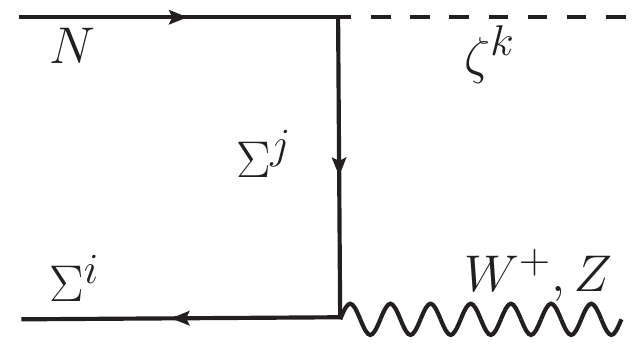}\,
        \includegraphics[scale=0.25]{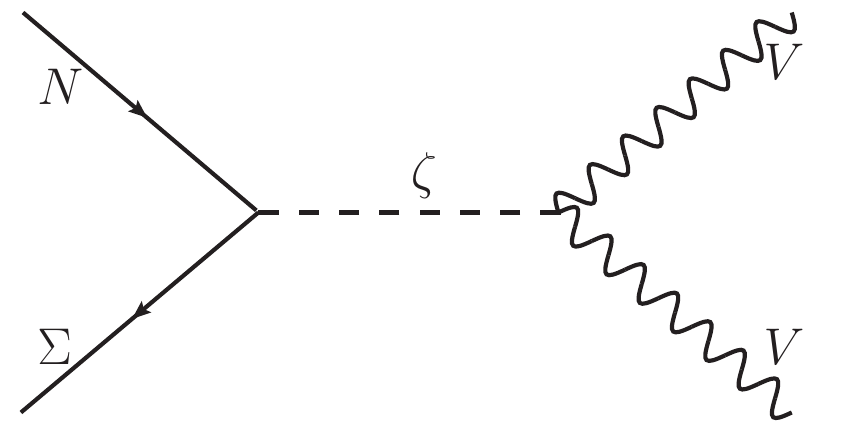}
    \caption{Dominant annihilation and co-annihilation channels of dark matter.}
    \label{fig:feynman}
\end{figure}

As mentioned above, the dark fermions $N$ and $\Sigma = (\Sigma^{++},\Sigma^+,\Sigma^0,\Sigma^-)$ are connected to the SM through $\zeta$ with the Yukawa coupling $y_N \zeta^\dagger \Sigma_L N_L + H.c.$ and the presence of an unbroken $\mathcal{Z}_2$ symmetry ensures the stability of either $N$ or the neutral component of $\Sigma$ to be a viable dark matter candidate of the model. As both $\zeta$ and $\Sigma$ are charged under SM gauge symmetry, they can be thermalized with the SM bath through their gauge interactions. 

Assuming $N$ is lighter than $\Sigma$, $N$ also can be thermally produced in the early universe for a sizable $y_N$ and eventually freezes out when its interaction rate drops below the expansion rate of the universe.  Assuming, $\zeta$ is lighter than the dark matter $N$, it can dominantly annihilate into the pair of $\zeta$ particles via the Yukawa interaction shown in equation \eqref{lag:tot}. There can also be significant contributions from the coannihilation with different components of $\Sigma$. The dominant annihilation and coannihilation channels of $N$ for different final states are shown in figure \ref{fig:feynman}. As a result, the important parameters which can affect the relic abundance are the dark matter mass $(M_{N})$, the Yukawa coupling $(y_N)$, and the mass splitting between $\Sigma$ and $N$ which is defined as $\Delta{M}=M_{\Sigma_i}-M_{N}$. 

\begin{figure}
    \centering
    \includegraphics[scale=0.34]{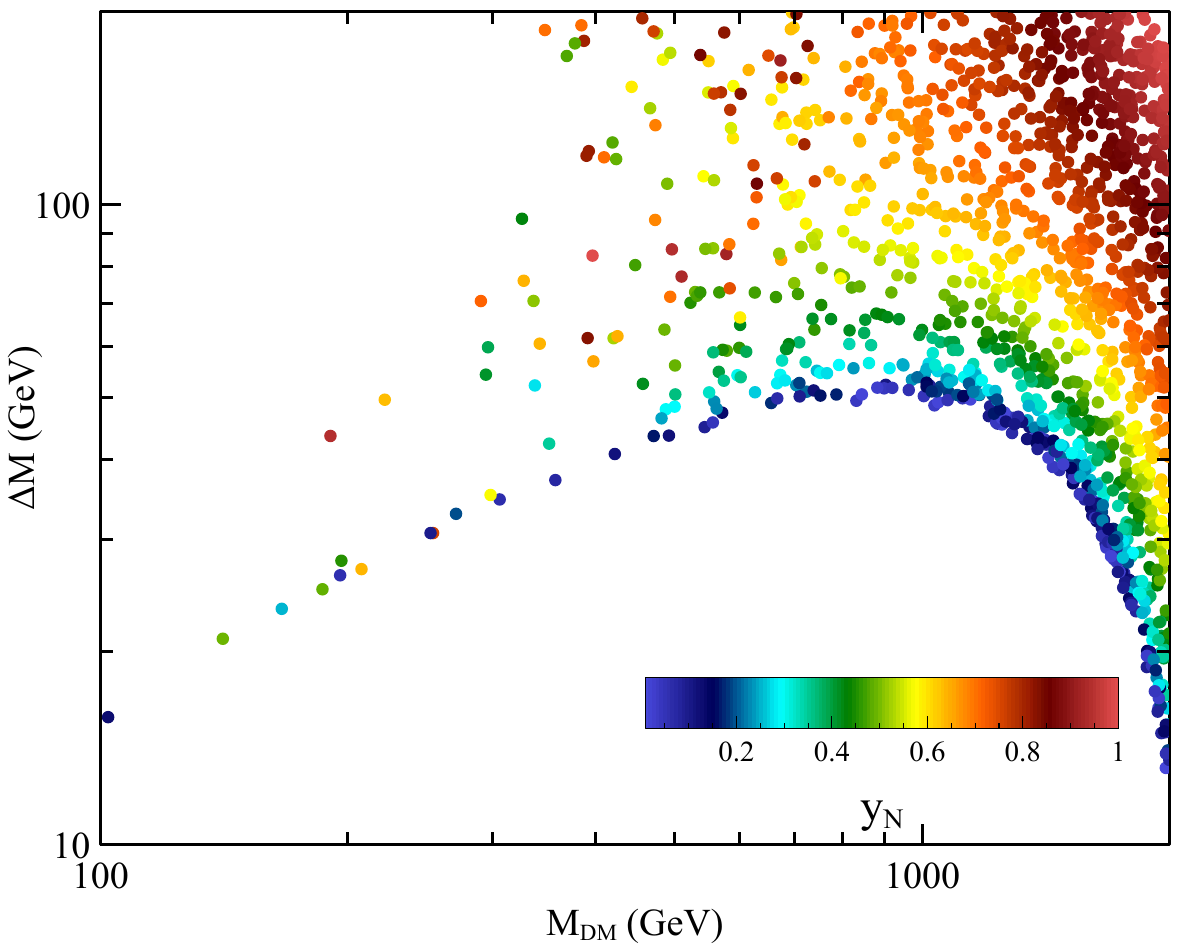}
    \caption{The allowed parameter space in $\Delta {M}-M_{N}$ plane from the relic density constraints where the colour code represents the variation of the Yukawa coupling $y_N$.}
    \label{fig:relic}
\end{figure}

In figure \ref{fig:relic}, we have shown the allowed parameter space in $M_{N}-\Delta{M}$ plane as the variation of $y_N$ is shown through the colour code. It is important to note that the relic density can only be satisfied for larger mass splitting $\Delta{M}$. For smaller mass splitting there can be huge coannihilation which suppresses the relic abundance which also sets a correlation with the Yukawa coupling $y_N$. The smaller the mass splitting stronger the coannihilation and one needs to reduce the Yukawa coupling to reduce the cross-section.

Another interesting dark matter phenomenology can arise in the case of tiny Yukawa coupling ($y_N$). This can lead to the non-thermal production of dark matter. For tiny $y_N$, dark matter can be directly produced from the direct decay of SM Higgs. The $3 \times 3$ mass matrix spanning $(N_L, \Sigma^0_L, \overline{\Sigma^0_R})$ is then

 \begin{equation}
{\cal M}_{N\Sigma} = \begin{pmatrix} m_N & y_N v_\zeta & 0 \cr y_N v_\zeta & 
0 & m_\Sigma \cr 0 & m_\Sigma & 0
\end{pmatrix}.
\end{equation}

Assuming that $m_N$ is of order GeV and $m_\Sigma$ is very much heavier, 
the $N-\Sigma$ mixing is $y_N v_\zeta/m_\Sigma$.
Consider now the decay of the SM Higgs $h$ to $NN$.  It does so first 
through $h-H$ mixing which is roughly $v_\zeta/3v_0$, then through $N-\Sigma$ 
mixing as just noted.  The effective coupling is 
\begin{equation}
f_h = \left( {v_\zeta \over 3 v_0} \right) \left( {y_N \over \sqrt{2}} \right) 
\left( {y_N v_\zeta \over m_\Sigma} \right).
\end{equation}
The decay rate of $h$ to $NN+\bar{N}\bar{N}$ is~\cite{Ma:2021eko}
\begin{equation}
\Gamma_h = {f_h^2 m_h \over 8\pi} \sqrt{1-4x^2}(1-2x^2),
\end{equation}
where $x=m_N/m_h$.  The correct dark matter relic abundance is 
obtained~\cite{Arcadi:2013aba} if $f_h \sim 10^{-12}x^{-1/2}$, provided that the 
reheat temperature of the universe is above $m_h$ but well below 
$m_H$ and $m_\Sigma$.  For $m_N \sim 1$ GeV, this implies 
$m_\Sigma/y_N^2 \sim (v_\zeta/{\rm GeV})^2 (1.2 \times 10^8~{\rm GeV})$.


We now turn to probe $\zeta^{\pm\pm}$ at the LHC. The sole method of producing the heavy Higgs Quadruplet $\zeta^{\pm\pm}$ at the LHC through proton-proton collisions is by generating a pair of $\zeta^{\pm \pm}$ particles. This can occur through three specific mediators: via SM Higgs boson, photon, and Z boson interactions, represented as $p p \to h/\gamma/z \to \zeta^{++} \zeta^{--}$. The combined cross-section for these processes amounts to $8.6 ~fb$. The $\zeta^{--}$ decay occurs through various channels, such as $\zeta^{--} \rightarrow W^- W^-, \zeta^{-} \zeta^{-}, W^- \zeta^{-}, \Sigma^{--} N$. In our benchmark points, the masses of $\Sigma^{--}$ and $\zeta^{-}$ exceed that of $\zeta^{--}$. Hence, the most favorable decay pathway for $\zeta^{--}$ is to $W^- W^-$, with a decay rate of $0.16 ~fb$. However, the exclusive decay of the doubly charged Higgs boson into same-sign $W$ bosons isn't exclusively tied to this specific model. It's also a characteristic feature of the Higgs triplet model within the Type-II seesaw framework \cite{Chiang:2012dk, Kanemura:2014goa, Melfo:2011nx,Ashanujjaman:2021txz}.

\begin{figure}
\centering
\includegraphics[scale=.2]{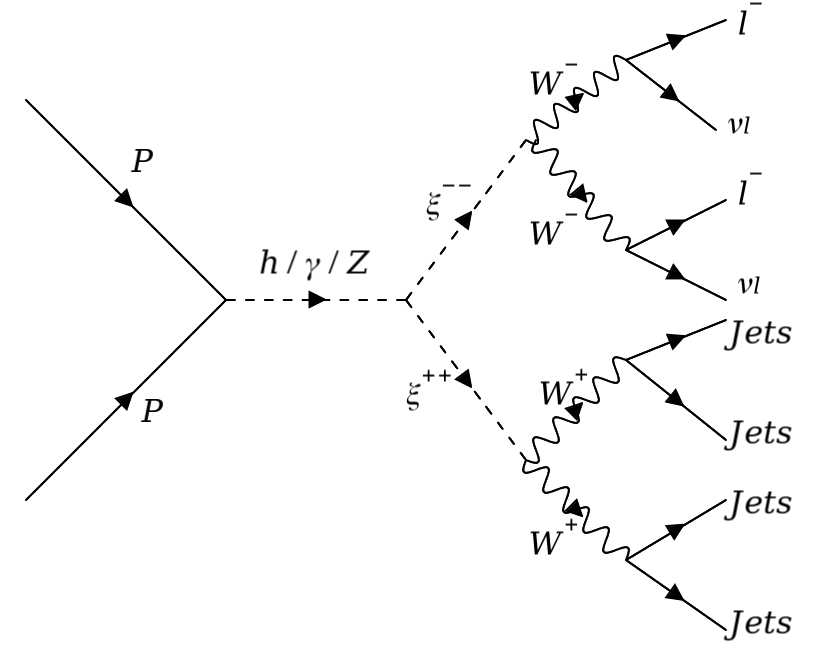}
\caption{\label{feynmanHhh} Feynman diagram for the process $p p \to \zeta^{++} \zeta^{--} \to 2W^+ 2W^- \to 4j 2l^- + MET$.}
\end{figure}
 
In our analysis, we consider an integrated luminosity of $L_{\text{int}}=3000~{\text{fb}}^{-1}$ at the center-of-mass energy of $\sqrt{s}=14~\text{TeV}$. To probe $\zeta^{\pm \pm}$, we explore the process $p p \to \zeta^{++} \zeta^{--} \to 2W^+ 2W^- \to 4j 2l^- +$ missing energy (MET), depicted by the Feynman diagram in figure ~\ref{feynmanHhh}. We employ two benchmark points: one with $m_{\zeta_{\pm\pm}}=400~\text{GeV}$ and a total cross section of $0.14 fb$, and another with $m_{\zeta_{\pm\pm}}=500~\text{GeV}$ and a total cross section of $0.05 fb$. To detect $\zeta^{\pm\pm}$ at the LHC, we examine the signal alongside the corresponding background arising from SM processes. The primary background processes interfering with our signal are: $pp\to WZ \to l^+ + 2l^- + \text{MET}$, $pp\to WWZ \to 2jets + l^+ +2l^- + \text{MET}$, and $pp\to WWW \to 2jets + 2l^- + \text{MET}$.

Our strategy to eliminate these backgrounds involves plotting various kinematic distributions for the signal versus the backgrounds. These include parameters like Missing Transverse Energy ($\cancel{E}_{T}$), Missing Transverse Energy ($\cancel{H}_{T}$), Scalar Sum of the Transverse Energy ($H_{T}$), Scalar Sum of the Transverse Energy of all Final-State Objects/Jets ($E_T/H_T$), Absolute Value of the Pseudo-Rapidity ($|\eta|$), Velocity ($\beta$), Transverse Energy ($E_{T_{l^- l^-}}$), Magnitude of the Three-Momentum ($P$), Angular Distance in the Transverse Plane between Objects ($\Delta_R$), among others. We strategically select cuts that effectively suppress the background while preserving our signal. Table ~\ref{cuttb600} presents the comprehensive list of all cuts employed to eliminate the backgrounds.

\begin{figure*}
\centering
\includegraphics[scale=.36]{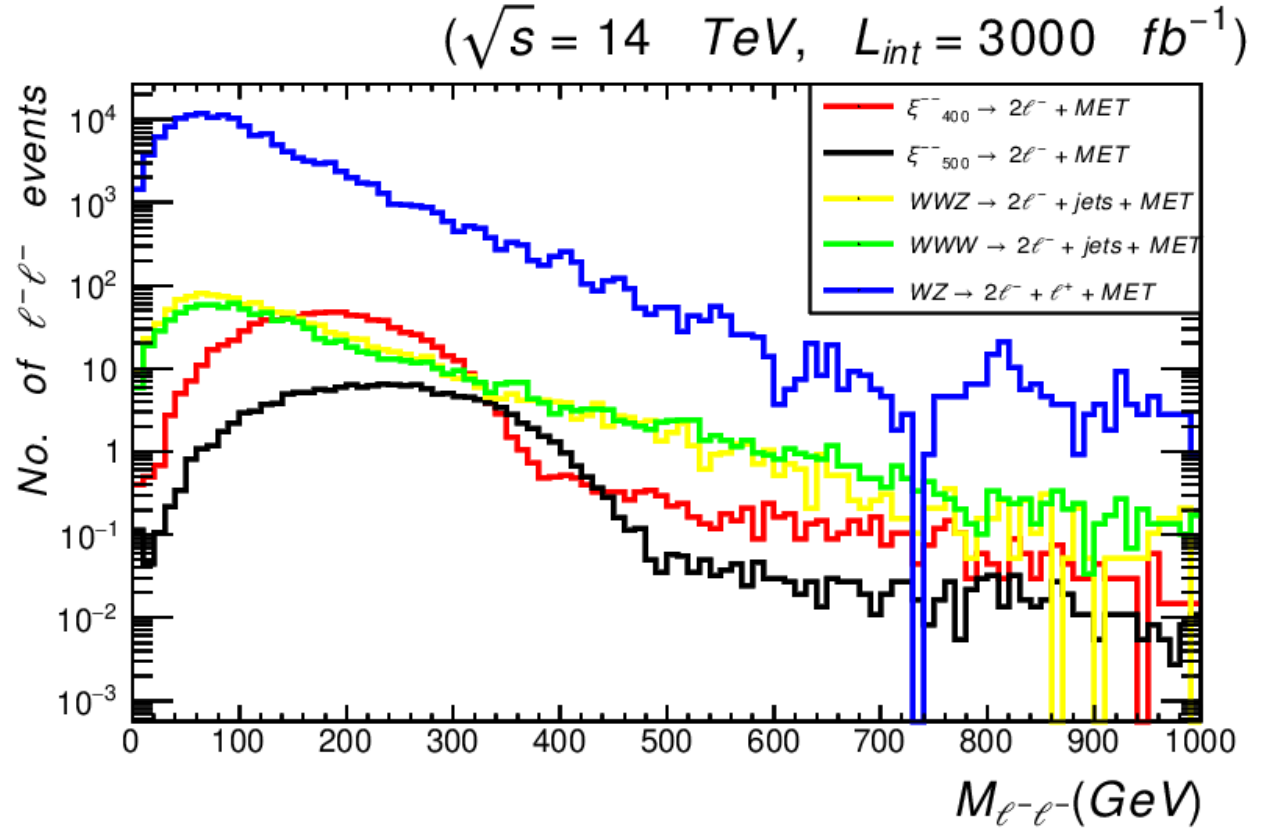}\,
\includegraphics[scale=.36]{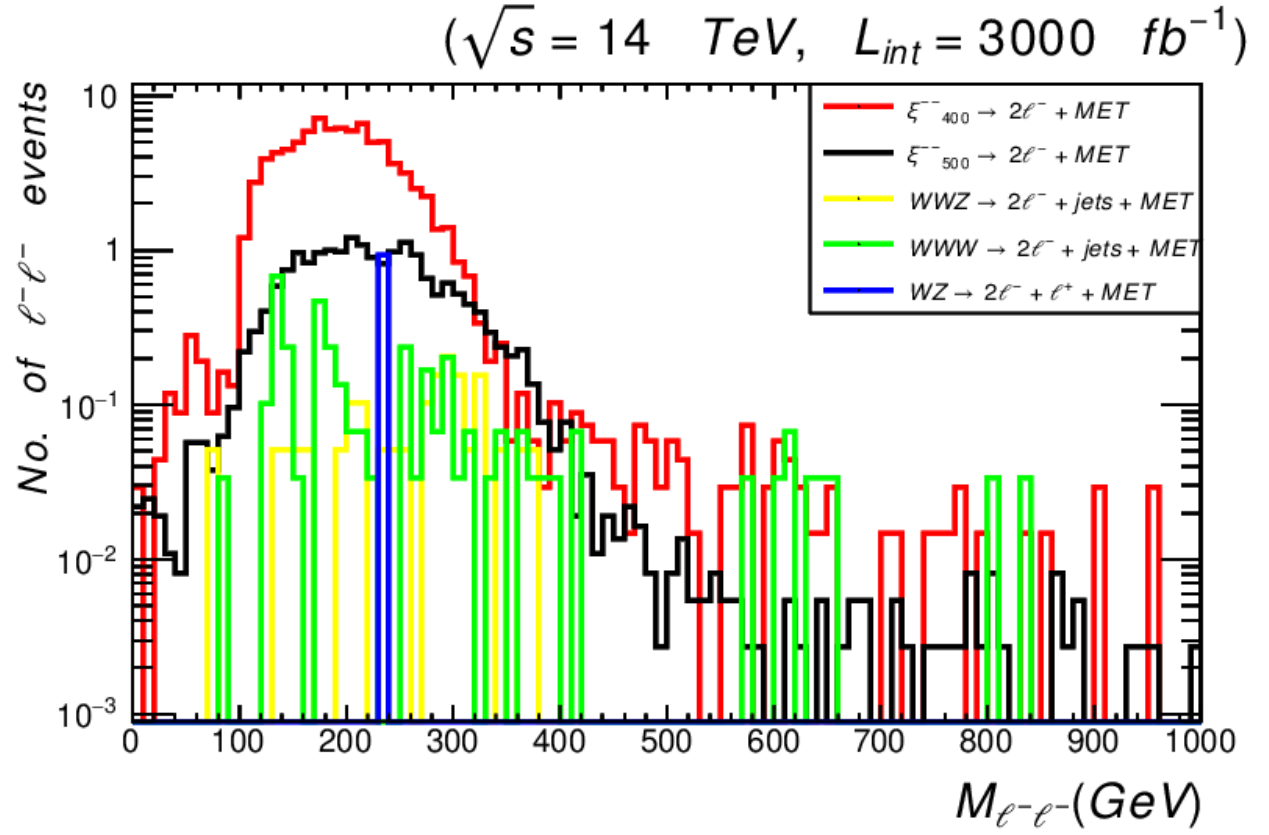}
\caption{\label{aftercuts}Number of signal events for the process at mass $m_{\zeta^{\pm\pm}}=400(500)~\text{GeV}$ (red(black)), alongside the relevant background events background (yellow/green/blue) before~(left) and after~(right).
}
\label{analysis}
\end{figure*}

The concluding distributions of these events are depicted in Fig~\ref{analysis}, showcasing the invariant mass distribution for two same-sign leptons both pre and post the application of cuts. A notable observation is that following the implementation of cuts, the signal exhibits a higher count of events compared to the background. We also explore another scenario to probe $\zeta^{\pm\pm}$ via the process $p p \to \zeta^{++} \zeta^{--} \to 2W^+ 2W^- \to 4l~(l=e,~\mu)+$ MET. However, in this scenario, the cross sections for masses of 400 and 500 GeV are notably small, amounting to $0.016~(0.0056)~\text{fb}$ for $m_{\zeta^{\pm\pm}}=400~(500)~\text{GeV}$. These extremely low cross sections (in fractions of fb) result in an exceedingly small number of events compared to the relevant background.

\begin{widetext}
\begin{center}
\begin{table}
\begin{tabular}{|c|c|c|c|}
\hline
Cuts (select)& Signal (S): $m_{\zeta^{\pm\pm}}=400(500)~\text{GeV}$ & Backgrounds ($\sum B$) & S/$\sqrt{\text{B}}$\\\hline
Initial (no cut) & $1620.0$ ($272.0$)  & $459213.0$ & $2.4$ ($0.4$)\\ \hline
$H_{T} > 250.0~\text{GeV}$ & $1619.8  $($272.0$) & $65102.0 $ & $6.4$($1.7$) \\ \hline
$\cancel{H}_T > 400.0~\text{GeV}$ & $632.1 $($143.0$) & $958.0 $ & $20.8$($4.7$)\\ \hline
$(\Delta R)_{l^- l^-} > 0.5$ & $393.9 $($85.4$) & $465.2$ & $18.2$($4.3$)\\ \hline
$\beta_{l^- l^-} > 0.75$ & $361.8 $($73.1$) & $232.4 $ & $23.8$($5.4$) \\\hline
$M_T > 100.0~\text{GeV}$ & $241.1 $($57.9$) & $102.6$ & $23.9$($5.9$) \\ \hline
$P_{l^- l^-} < 550.0~\text{GeV}$ & $202.4 $($34.0$) & $55.6 $ & $27.6$($4.6$) \\ \hline
$\phi_{l^- l^-} > 0.0$ & $101.9 $($17.4$) & $5.1 $ & $44.7$($7.7$) \\ \hline
\end{tabular}
\caption{\label{cuttb600}Cut flow charts for the signal versus its relevant background and the corresponding number of events and significance.}
\end{table}
\end{center}
\end{widetext}

\onecolumngrid
\onecolumngrid

The work of K. E. and S.K. is partially supported by Science, Technology $\&$ Innovation Funding Authority (STDF) under grant number 48173. The work of EM is supported in part by the U.$~$S.$~$ Department of Energy Grant 
No. DE-SC0008541. The work of DN is supported by the National Research Foundation of Korea (NRF)’s grants, grant no. 2019R1A2C3005009(DN). DN also acknowledges Debasish Borah and Sanjoy Mandal for useful discussions.

\bibliographystyle{apsrev4-1}
\bibliography{ref}

\end{document}